\documentclass[twocolumn,prl,aps]{revtex4}

\usepackage{epsfig}
\usepackage{axodraw}

\newcommand{\bed}{\begin{displaymath}}
\newcommand{\eed}{\end{displaymath}}
\newcommand{\beq}{\begin{equation}}
\newcommand{\eeq}{\end{equation}}
\newcommand{\bea}{\begin{eqnarray}}
\newcommand{\eea}{\end{eqnarray}}
\newcommand{\tgb}{{\rm tg}\beta}

\newcommand{\sq}{\tilde{q}}
\newcommand{\st}{\tilde{t}}
\newcommand{\sgl}{\tilde{g}}

\newcommand{\lsim}{\raisebox{-0.13cm}{~\shortstack{$<$ \\[-0.07cm] $\sim$}}~}

\begin{document}

\title{
\begin{flushright} \begin{small} {\rm TTP13--029} \\ {\rm
SFB/CPP--13--63} \\ {\rm PSI--PR--13--12} \\ \end{small} \end{flushright}
MSSM Higgs Self-Couplings: Two-Loop ${\cal O}(\alpha_t\alpha_s)$ Corrections}

\author{Mathias Brucherseifer}
\affiliation{Institut f\"ur Theoretische Teilchenphysik, KIT, D--76128
Karlsruhe, Germany}

\author{Ryan Gavin}
\affiliation{Paul Scherrer Institut, CH--5232 Villigen PSI, Switzerland}

\author{Michael Spira}
\affiliation{Paul Scherrer Institut, CH--5232 Villigen PSI, Switzerland}


\begin{abstract} 
\noindent
We analyze the minimal supersymmetric Higgs self-couplings at ${\cal
O}(\alpha_t\alpha_s)$ within the effective potential approach. The
two-loop corrections turn out to be of moderate size in the
$\overline{\rm DR}$ scheme if the central scale is chosen as half the
SUSY scale. The inclusion of the two-loop corrections reduces the
renormalization scale dependence to the per-cent level. These
results have a significant impact on measurements of the
trilinear Higgs self-couplings at the LHC and a future $e^+e^-$
collider.
\end{abstract}


\maketitle

\noindent
The Higgs mechanism~\cite{Higgs:1964ia} is a cornerstone of the Standard
Model (SM) and its supersymmetric extensions. The masses of the
fundamental particles, i.e.~electroweak gauge bosons, leptons and
quarks, are generated by interactions with Higgs fields. The recently
discovered particle with a mass of $\sim$125 GeV at the LHC
\cite{higgsdiscovery} seems to be the SM Higgs boson, i.e.~all
tested properties such as its couplings to the other SM particles, its spin
and ${\cal CP}$ quantum numbers agree with the SM predictions
\cite{hcoup}. However, the errors of the measured couplings to fermions
and vector bosons leave room for deviations from the SM values, which
naturally arise in SM extensions as e.g.~the minimal supersymmetric
extension (MSSM).

The MSSM requires the introduction of two Higgs doublets. After
electroweak symmetry breaking there are five elementary Higgs particles,
two ${\cal CP}$-even ($h,H$), one ${\cal CP}$-odd ($A$) and two charged
($H^\pm$). At lowest order all couplings and masses of the MSSM
Higgs sector are fixed by two independent input parameters, which are
generally chosen as $\tgb=v_2/v_1$, the ratio of the two vacuum
expectation values (vevs) $v_{1,2}$, and the pseudoscalar Higgs mass
$M_A$. Including the one-loop and dominant two- and three-loop
corrections the upper bound on the light scalar Higgs mass is $M_h\lsim
135$ GeV \cite{mssmrad}. The Higgs boson couplings to fermions and gauge
bosons depend on mixing angles $\alpha$ and $\beta$, which are defined
by diagonalizing the neutral and charged Higgs mass matrices.

One of the most important tests of the Higgs sector in the future is the
measurement of the Higgs potential, i.e.~the self-interactions of the
Higgs particles. It is possible that the trilinear Higgs self-coupling
could be measured at the LHC after the high-luminosity upgrade
\cite{lhcself}, while a measurement of the quartic Higgs coupling will
be out of reach at any foreseen collider due to the tiny signal rates
\cite{quartic}.  Within the MSSM the prospects for the trilinear Higgs
couplings can be better due to the possible appearance of resonant Higgs
decays into lighter Higgs pairs as e.g.~the heavy scalar Higgs decay in
$gg\to H\to hh$ for values of $\tgb\lsim 10$ \cite{gg2hh}. The proper
treatment of the signal rates within the MSSM requires the determination
of the radiative corrections to the effective trilinear Higgs couplings
supplemented by moderate process-dependent corrections \cite{htohh}.
Many years ago the one-loop corrections to the effective trilinear Higgs
couplings have been shown to be large \cite{tril1loop,tril1loop2}.
However, sizable residual uncertainties of these effective couplings,
arising by integrating out the heavy SUSY particles and the top quark,
are left over. In order to reduce these uncertainties a two-loop
calculation of the trilinear Higgs couplings is required. The one-loop
corrections are dominated entirely by top and stop loop contributions.
Only for large values of $\tgb$ can the bottom/sbottom loop
contributions become relevant thanks to the large enhancement of the
bottom Yukawa couplings in this regime \cite{tril1loop}. In this work we
will describe the two-loop SUSY--QCD corrections to the top/stop-loop
induced corrections \cite{mathias}.

We will parametrize the two MSSM Higgs doublets as
\bea
H_1 = \left(\begin{array}{c} H_1^0 \\ -H_1^- \end{array} \right), \qquad
H_2 = \left(\begin{array}{c} H_2^+ \\ H_2^0 \end{array} \right) \, ,
\eea
where $H_{1,2}^\pm$ denote the four charged Higgs fields in the current
eigenstate basis. The neutral Higgs fields decompose into $v_{1,2}$ and
scalar/pseudoscalar components as
\bea
H_j^0 & = & \frac{1}{\sqrt{2}} (v_j + S_j + i P_j) \quad (j=1,2) \; .
\eea
The neutral physical Higgs and would-be Goldstone fields emerge from
rotations by the mixing angles $\alpha$ and $\beta$,
\bea
S_1 & = & H c_\alpha - h s_\alpha, \qquad \quad
P_1   =   G^0 c_\beta - A s_\beta, \nonumber \\
S_2 & = & H s_\alpha + h c_\alpha, \qquad \quad
P_2   =   G^0 s_\beta + A c_\beta \; .
\eea
The vevs are defined as $v_1 = v c_\beta, v_2 = v s_\beta$ with
$v\approx 246$ GeV.  The tree-level Higgs potential is given by
\bea
V_0 & = & m_1^2 |H_1|^2 + m_2^2 |H_2|^2 - B\mu\epsilon_{ij} (H_1^iH_2^j
+ h.c.) \nonumber \\
& + & \frac{g^2+g'^2}{8} (|H_1|^2-|H_2|^2)^2 + \frac{g^2}{2}
|H_1^\dagger H_2|^2 \; ,
\eea
where $m_{1,2}^2=m_{H_{1,2}}^2 + \mu^2$ with $m_{H_{1,2}},\mu$
denoting the soft SUSY-breaking Higgs and Higgsino mass parameters
respectively. The object $\epsilon_{ij}$ is the antisymmetric
two-dimensional tensor, while $g,g'$ are the isospin and hypercharge
gauge couplings. The parameters $m_{1,2}$ are eliminated by the
minimization condition of the effective potential, while the parameter
$B\mu$ is traded for the pseudoscalar mass $M_A$. Taking the second
derivatives of $V_0$ with respect to the Higgs fields yields the mass
matrices, while the third and fourth derivatives define the trilinear
and quartic Higgs couplings, respectively. After rotation to the
physical mass eigenstates, the neutral trilinear Higgs couplings at
leading order are given by
\bea
\lambda_{hhh} & = & 3\frac{M_Z^2}{v}c_{2\alpha}s_{\alpha+\beta}, \quad
\lambda_{HHH} = 3\frac{M_Z^2}{v}c_{2\alpha}c_{\alpha+\beta}, \nonumber \\
\lambda_{Hhh} & = & \frac{M_Z^2}{v} [2s_{2\alpha}s_{\alpha+\beta} -
c_{2\alpha}c_{\alpha+\beta}], \nonumber \\
\lambda_{HHh} & = & -\frac{M_Z^2}{v} [2s_{2\alpha}c_{\alpha+\beta} +
c_{2\alpha}s_{\alpha+\beta}], \nonumber \\
\lambda_{hAA} & = & \frac{M_Z^2}{v}c_{2\beta}s_{\alpha+\beta}, \quad
\lambda_{HAA} = -\frac{M_Z^2}{v}c_{2\beta}c_{\alpha+\beta} \; .
\eea
\noindent
{\bf One-Loop Corrections.} These couplings are subject to radiative
corrections. Using dimensional reduction in $n=4-2\epsilon$ dimensions,
the leading top/stop-induced corrections of ${\cal O}(\alpha_t)$ to the
effective potential in Landau gauge can be cast into the form
\cite{veff1}
\bea
V_1 & = & \frac{3}{16\pi^2} \left\{ \bar m_t^4 \left[
\frac{1}{\epsilon} + \frac{3}{2} - \log\frac{\bar m_t^2}{Q^2} \right.
\right. \nonumber \\
& & \qquad \quad \left. + \epsilon\left(\frac{7}{4} - \frac{3}{2}
\log\frac{\bar m_t^2}{Q^2} + \frac{1}{2} \log^2\frac{\bar m_t^2}{Q^2} +
\frac{1}{2}\zeta_2\right)\right] \nonumber \\
& & \qquad \quad \left. - \frac{1}{2}[(\bar m_t \leftrightarrow \bar
m_{\tilde t_1}) + (\bar m_t \leftrightarrow \bar m_{\tilde t_2}) ]
\right\}
\eea
with the field-dependent mass parameters defined as
\bea
\bar m_t^2 & = & |X|^2 \; , \nonumber \\
\bar m_{\tilde t_{1,2}}^2 & = & \frac{1}{2} \left( \tilde M_{\tilde t_R}^2 +
\tilde M_{\tilde t_L}^2 + 2\bar m_t^2 \right. \nonumber \\
& & \left. \mp \sqrt{(\tilde M_{\tilde t_L}^2-\tilde M_{\tilde t_R}^2)^2 + 4|\tilde X|^2} \right)
\; , \nonumber \\
X & = & h_t H_2^0 , \qquad
\tilde X = h_t \left[ A_t H_2^0 - \mu H_1^{0*} \right] \; ,
\eea
where the parameters $\tilde{M}_{\tilde t_{L/R}}$ include the $D$-terms,
\bea
& & \!\!\!\!\!\!\!\! \tilde M^2_{\tilde t_{L/R}} = M^2_{\tilde t_{L/R}}
+ D_{\tilde t_{L/R}} \; , \nonumber \\
& & \!\!\!\!\!\!\!\! D_{\tilde t_L} = M_Z^2 \left(\frac{1}{2} -
\frac{2}{3} s^2_{W} \right) c_{2\beta} , \; D_{\tilde t_R} =
M_Z^2~\frac{2}{3} s^2_{W} c_{2\beta} \; .
\eea
The scale $Q^2$ is related to the 't Hooft mass scale $\mu_0$ as $Q^2 =
4\pi \mu_0 e^{-\gamma_E}$ with the Euler-constant $\gamma_E$. The top
Yukawa coupling $h_t = \sqrt{2}m_t/(v s_\beta)$ defines the
corresponding coupling $\alpha_t=h_t^2/(4\pi)$.  Including the
loop-corrected minimization condition and the loop-corrected
pseudoscalar Higgs mass the third derivative of this effective potential
reproduces the results of Ref.~\cite{tril1loop} for the trilinear Higgs
couplings. Analogously the quartic Higgs couplings can be derived from
the fourth derivatives.

\noindent
{\bf Two-Loop Corrections.} The two-loop SUSY--QCD corrections to the
effective potential are given by \cite{veff2}
\bea
V_2 \!\! & = & \frac{\alpha_s}{8\pi^3}\Big\{ J(\bar m_t^2,\bar m_t^2) -
2\bar m_t^2 I(\bar m_t^2,\bar m_t^2,0) \nonumber \\
& + & \left[\frac{1}{4}(2-s_{2\bar\theta}^2) J(\bar
m_{\tilde t_1}^2,\bar m_{\tilde t_1}^2) + \frac{s_{2\bar\theta}^2}{4}
J(\bar m_{\tilde t_1}^2,\bar m_{\tilde t_2}^2) \right. \nonumber \\
& + & \bar m_{\tilde t_1}^2
I(\bar m_{\tilde t_1}^2,\bar m_{\tilde t_1}^2,0)
+ J(m_{\tilde g}^2,\bar m_t^2)
- J(\bar m_{\tilde t_1}^2,m_{\tilde g}^2) \nonumber \\
& - & J(\bar m_{\tilde t_1}^2,\bar m_t^2)
- (\bar m_{\tilde t_1}^2-m_{\tilde g}^2-\bar m_t^2)
I(\bar m_{\tilde t_1}^2,m_{\tilde g}^2,\bar m_t^2)
\nonumber \\
& - & \!\! \left. \left. \!\!\!\! 2 m_{\tilde g} \bar\xi
I(\bar m_{\tilde t_1}^2,m_{\tilde g}^2,\bar m_t^2) + (\bar m_{\tilde
t_1}^2 \leftrightarrow \bar m_{\tilde t_2}^2, \bar\xi \to -\bar\xi )
\right] \right\}
\eea
with the additional field-dependent parameters
\bea
s_{2\bar\theta}^2 & = & \frac{4|\tilde X|^2}{(m_{\tilde t_1}^2-m_{\tilde
t_2}^2)^2} \; , \nonumber \\
\bar \xi & = & 2\frac{\Re e (X) \Re e(\tilde X) + \Im m (X) \Im m(\tilde
X)}{m_{\tilde t_1}^2-m_{\tilde t_2}^2} \; .
\eea
The gluino mass is denoted by $m_{\tilde g}$ and the two-loop integrals
$I,J$ are defined and calculated in \cite{t134}.  We have calculated the
derivatives of the two-loop corrected Higgs potential up to fourth order
in the Higgs fields. After implementing the minimization condition and
the pseudoscalar Higgs mass at the two-loop level, we have renormalized
the top mass, stop masses, stop mixing angle and $A_t$ parameters of the
one-loop corrected Higgs potential within the $\overline{\rm
DR}$ scheme. This scheme choice ensures the relation between the stop
mixing angle and $A_t$,
\beq
s_{2\theta} = \frac{2m_t(A_t-\mu / \tgb)}{m_{\tilde t_1}^2-m_{\tilde
t_2}^2} \; .
\label{eq:mixang}
\eeq
The $\overline{\rm DR}$ counter terms are given by
\bea
\delta m_t & = & -C_F \frac{\alpha_s}{2\pi} m_t \left[
\frac{1}{\epsilon}+\log\frac{Q^2}{\mu_R^2} \right]\; , \nonumber \\
\delta m_{\tilde t_{1/2}}^2 & = & C_F \frac{\alpha_s}{4\pi} \left[ \mp
(m_{\tilde t_1}^2-m_{\tilde t_2}^2) s^2_{2\theta} \right. \nonumber \\
& & -4(m_{\tilde g}^2 + m_t^2 \mp m_{\tilde g} m_t s_{2\theta}) \Big]
\left[\frac{1}{\epsilon}+\log\frac{Q^2}{\mu_R^2} \right] \; , \nonumber \\
\delta A_t & = & C_F \frac{\alpha_s}{\pi} m_{\tilde g}
\left[\frac{1}{\epsilon}+\log\frac{Q^2}{\mu_R^2} \right] \; .
\eea
The $\overline{\rm DR}$ counter term $\delta\theta$ can be derived from
the relation (\ref{eq:mixang}).  After renormalization we reproduce the
known two-loop results for the Higgs masses \cite{veff2,slavich} and
arrive at finite expressions for the trilinear and quartic Higgs
couplings. These are finally rotated to the physical Higgs mass
eigenstates by the radiatively corrected mixing angles $\alpha,\beta$,
which diagonalize the two-loop corrected scalar and pseudoscalar Higgs
mass matrices. At two-loop order we have checked explicitly that the
trilinear and quartic Higgs couplings $\lambda_{hhh},\lambda_{hhhh}$
approach their two-loop SM limits for large $M_A$ and SUSY masses in
analogy with the one-loop analysis of Ref.~\cite{tril1loop2}.

In order to obtain consistent results for the Higgs self-couplings we
used the following expressions for the running $\overline{\rm DR}$
parameters at the renormalization scale $\mu_R$ \cite{hsqsq},
\bea
m_t (\mu_R) & = & m_t (M_t)
\left(\frac{\alpha_{s}(\mu_R)}{\alpha_{s}(M_t)}\right)^{\frac{8}{9}}
\frac{9+10\alpha_{s}(\mu_R)/\pi}{9+10\alpha_{s}(M_t)/\pi} \; ,
\nonumber \\
m_{\tilde g}(\mu_R) & = & m_{\tilde
g}(M_{\tilde g})~\frac{\alpha_{s}(\mu_R)~[6-7\alpha_{s}(\mu_R)/\pi]}
{\alpha_{s}(M_{\tilde g})~[6-7\alpha_{s}(M_{\tilde g})/\pi]} \; , \nonumber \\
A_t(\mu_R) & = & A_t(Q_0) + m_{\tilde g}(Q_0) \left\{-\frac{16}{9}\,
\left[\frac{\alpha_{s}(\mu_R)}{\alpha_{s}(Q_0)}-1\right] \times \right.
\nonumber \\
& & \left. \!\!\!\!\!\!\!\!\!\!\!\!
\left[1+\frac{7}{6}\,\frac{\alpha_{s}(Q_0)}{\pi}
\right] - \frac{4}{27}\,\frac{\alpha_{s}(Q_0)}{\pi}
\left[\frac{\alpha_{s}^2(\mu_R)}{\alpha_{s}^2(Q_0)}-1\right]
\right\} \; , \nonumber \\
M_{\!\tilde t_{\!L\!/\!R}}^2 \! (\mu_R) \!\!\! & = &
M_{\tilde t_{L\!/\!R}}^2(Q_0)
+ m_{\tilde g}^2(Q_0) \left\{\frac{8}{9}\,
\left[\frac{\alpha_{s}^2(\mu_R)}{\alpha_{s}^2(Q_0)}-1\right]
\times \right. \nonumber \\
& & \left. \!\!\!\!\!\!\!\!\!\!\!
\left[1+\frac{7}{3}\,\frac{\alpha_{s}(Q_0)}{\pi}
\right] - \frac{8}{81}\,\frac{\alpha_{s}(Q_0)}{\pi}
\left[\frac{\alpha_{s}^3(\mu_R)}{\alpha_{s}^3(Q_0)}-1\right]
\right\} \; , \nonumber \\
\alpha_{s}(\mu_R) & = & \!\!\! \frac{4\pi}{\displaystyle
3\log (\mu_R^2/\Lambda^2)} \left\{ 1+\frac{14}{9}\,
\frac{\displaystyle \log\log (\mu_R^2/\Lambda^2)}
{\displaystyle \log (\mu_R^2/\Lambda^2)} \right\} ,
\label{eq:run}
\eea
where $Q_0$ denotes the input scale for these parameters and $\Lambda$
the QCD scale of the strong coupling $\alpha_s$ \footnote{Note that due
to a sign typo the formula for the running coupling $\alpha_s$ of
Ref.~\cite{hsqsq} differs from Eq.~(\ref{eq:run}).}. These expressions
are valid up to the next-to-leading-log level of the renormalization
group equations \cite{rge}. The $\overline{\rm DR}$ stop masses
$m_{\tilde t_i}(\mu_R)$ are obtained from the running SUSY-breaking
parameters as [$\tilde{M}_{\tilde t_{L/R}} = \tilde{M}_{\tilde
t_{L/R}}(\mu_R)$]
\bea
m_{\tilde t_{1/2}}^2(\mu_R) = m_t^2(\mu_R) +
\frac{1}{2}\left[ \tilde{M}_{\tilde t_L}^2 + \tilde{M}_{\tilde
t_R}^2
\right. \qquad \qquad \quad \nonumber \\
\left. \mp \sqrt{[\tilde{M}_{\tilde t_L}^2
- \tilde{M}_{\tilde t_R}^2]^2 + 4 m_t^2(\mu_R) [A_t(\mu_R) -
\mu/\tgb]^2} \right] .
\eea
Our running $\overline{\rm DR}$ parameters include the contributions of
all strongly interacting SM and SUSY particles. The top mass $m_t$, the
gluino mass $m_{\tilde g}$ and the strong coupling $\alpha_s$ are
related to the top pole mass $M_t$, the gluino pole mass $M_{\tilde g}$
and the 5-flavor $\overline{\rm MS}$ coupling $\alpha_{s,\overline{\rm
MS}}^{(5)}$ by
\bea
m_t(M_t) & = & M_t \left\{1+
\frac{\alpha_{s,\overline{\rm MS}}^{(5)}(M_t)}{3\pi}\Big[ 5 \right.
\nonumber \\
& & \quad + \sum_{i=1}^2 \Big( \bar
B_1(M_t^2;M_{\sgl},M_{\st_i},M_t^2) \nonumber \\
& & \left. \left. \left. - (-1)^i \frac{M_{\sgl}s_{2\theta}}{M_t} \bar
B_0(M_t^2;M_{\sgl},M_{\st_i},M_t^2) \right) \right] \right\}^{-1}
\!\!\!\! , \nonumber \\
\!\! m_{\tilde g}(M_{\tilde g}) \!\! & = & M_{\sgl} \left\{
1+\frac{\alpha_{s,\overline{\rm MS}}^{(5)}(M_{\sgl})}{4\pi}
\Big[ 15 \right. \nonumber \\
& & \quad + \sum_{q,i} \Big( \bar B_1(M_{\sgl}^2;M_q,M_{\sq_i},M_{\tilde
g}^2) \nonumber \\
& & \left. \left. \left. \!\!\!\! -(-1)^i \frac{M_q s_{2\theta_q}}{M_{\sgl}}
\bar B_0(M_{\sgl}^2;M_q,M_{\sq_i},M_{\tilde g}^2) \right) \right]
\right\}^{-1} \!\!\!\! , \nonumber \\
\alpha_{s}(Q_0) & = & \alpha_{s,\overline{\rm MS}}^{(5)}(Q_0) \left\{ 1 +
\frac{\alpha_{s,\overline{\rm MS}}^{(5)}(Q_0)}{\pi}\left[
\frac{1}{6}\log\frac{Q_0^2}{M_t^2} \right. \right. \nonumber \\
& & \left. \left. + \frac{1}{2}\log\frac{Q_0^2}{M_{\sgl}^2} + \frac{1}{24}
\sum_{\tilde q_i} \log\frac{Q_0^2}{M_{\tilde q_i}^2} + \frac{1}{4} \right]
\right\} \; ,
\eea
where $M_{\tilde q_i}$ denotes the squark pole mass and $\theta_q$ the
corresponding mixing angle generically (which have been defined via the
tree level relations). The finite parts of the one-loop integrals can be
cast into the form \cite{oneloop}
\bea
\bar B_{0[1]}(p^2;m_1,m_2,Q^2) = \Re e \int_0^1 dx~[-x] \times \qquad \qquad
\nonumber \\
\log\frac{Q^2}{m_1^2x + m_2^2(1-x)-p^2 x(1-x) - i \epsilon} \; ,
\eea
where the factor $-x$ has to be inserted for $\bar B_1$.

The numerical analysis of the Higgs self-couplings is performed in
the ``$m_h^{mod+}$'' MSSM scenario \cite{bench} as a representative case:
\bea
\tgb & = & 5,\;
M_{\tilde Q_{L/R}} = 1~{\rm TeV},\;
M_{\sgl} = 1.5~{\rm TeV}, \nonumber \\
A_b & = & A_t = 1.64~{\rm TeV},\;
\mu = 200~{\rm GeV} \; ,
\eea
where the parameters $M_{\tilde Q_{L/R}},A_t,A_b$ are defined at the
input scale $Q_0=M_{\tilde Q_{L/R}}$.  Within this scenario resonant
Higgs production $gg\to H\to hh$ occurs with a sizable cross section.
For the Higgs masses and couplings we used our calculation based on the
${\cal O}(\alpha_t\alpha_s)$-corrected effective potential. The top
quark pole mass has been chosen as $M_t=173.2$ GeV, while the strong
coupling constant has been normalized to $\alpha_{s,\overline{\rm
MS}}^{(5)}(M_Z)=0.119$.

\begin{figure}[t]
\begin{picture}(200,275)(0,0)
\put(30,140){\epsfig{file=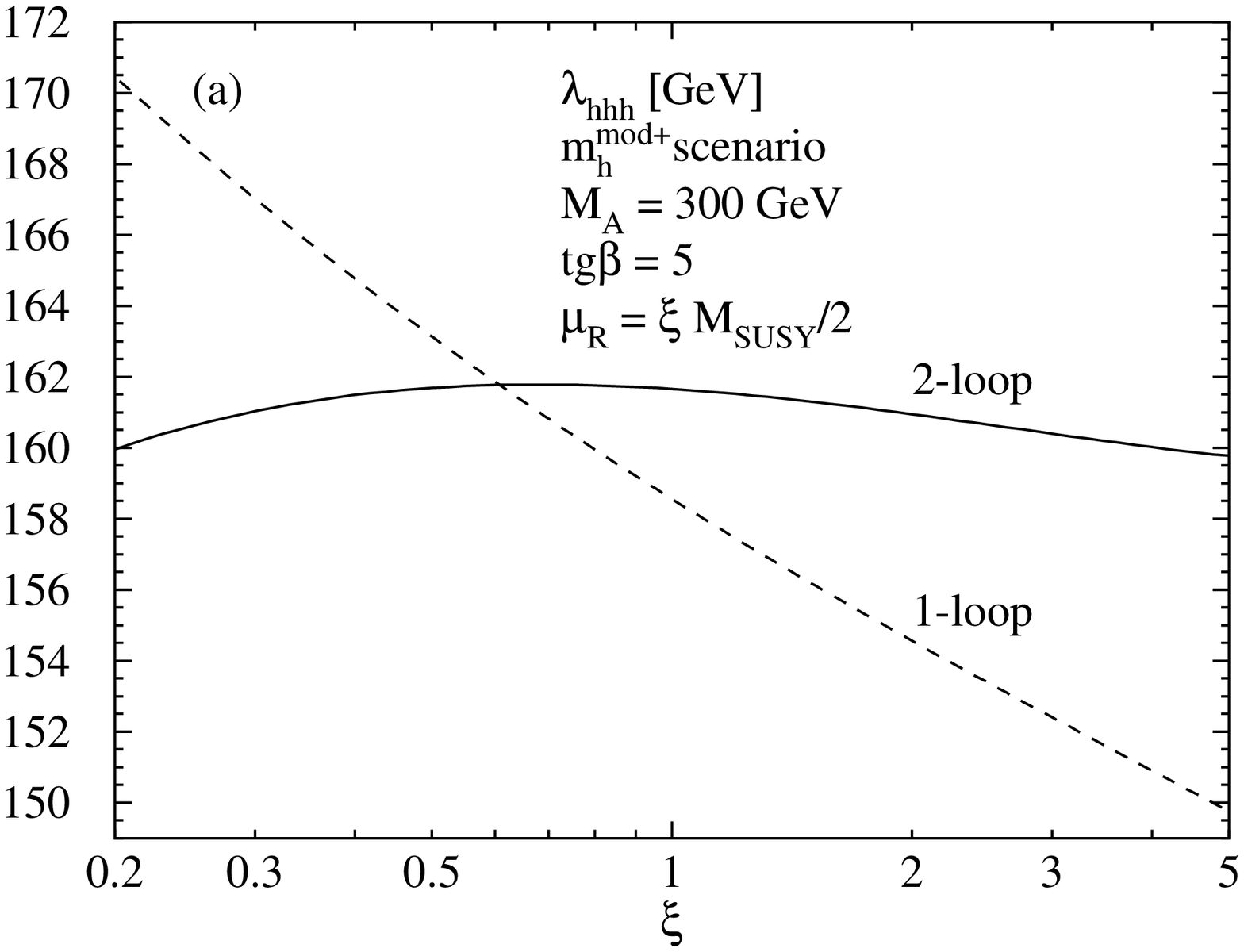,%
        bbllx=150pt,bblly=230pt,bburx=450pt,bbury=650pt,%
        scale=0.35,clip=}}
\put(30,0){\epsfig{file=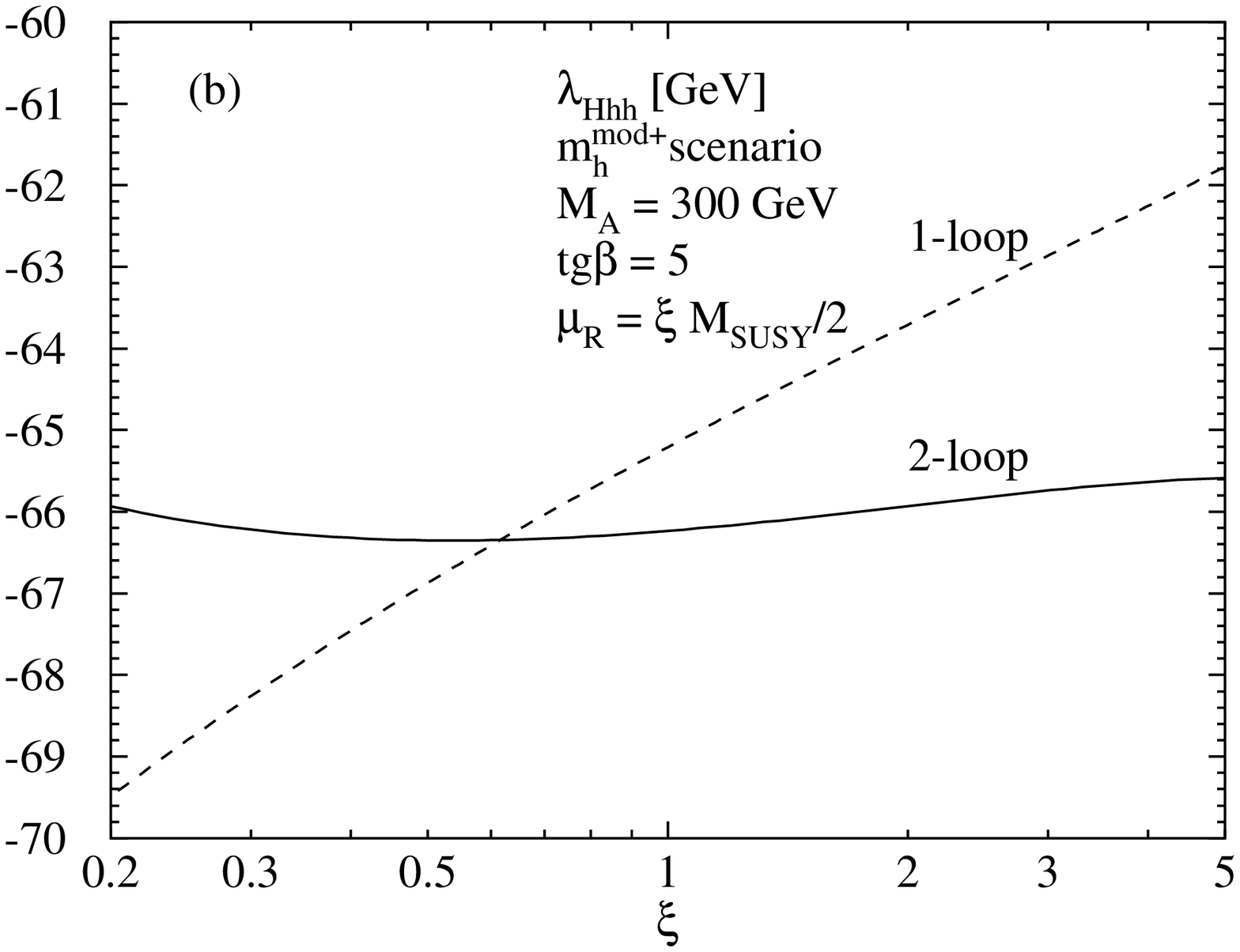,%
        bbllx=150pt,bblly=230pt,bburx=450pt,bbury=650pt,%
        scale=0.35,clip=}}
\end{picture}
\vspace*{-0.3cm}

\caption{\it \label{fg:scale} Scale dependence of the trilinear Higgs
couplings $\lambda_{hhh}$ (a) and $\lambda_{Hhh}$ (b) at one- and
two-loop order in the $m_h^{mod+}$ scenario for $M_A=300$ GeV and
$\tgb=5$.}
\vspace*{-0.60cm}

\end{figure}
\begin{figure}[t]
\begin{picture}(200,275)(0,0)
\put(30,140){\epsfig{file=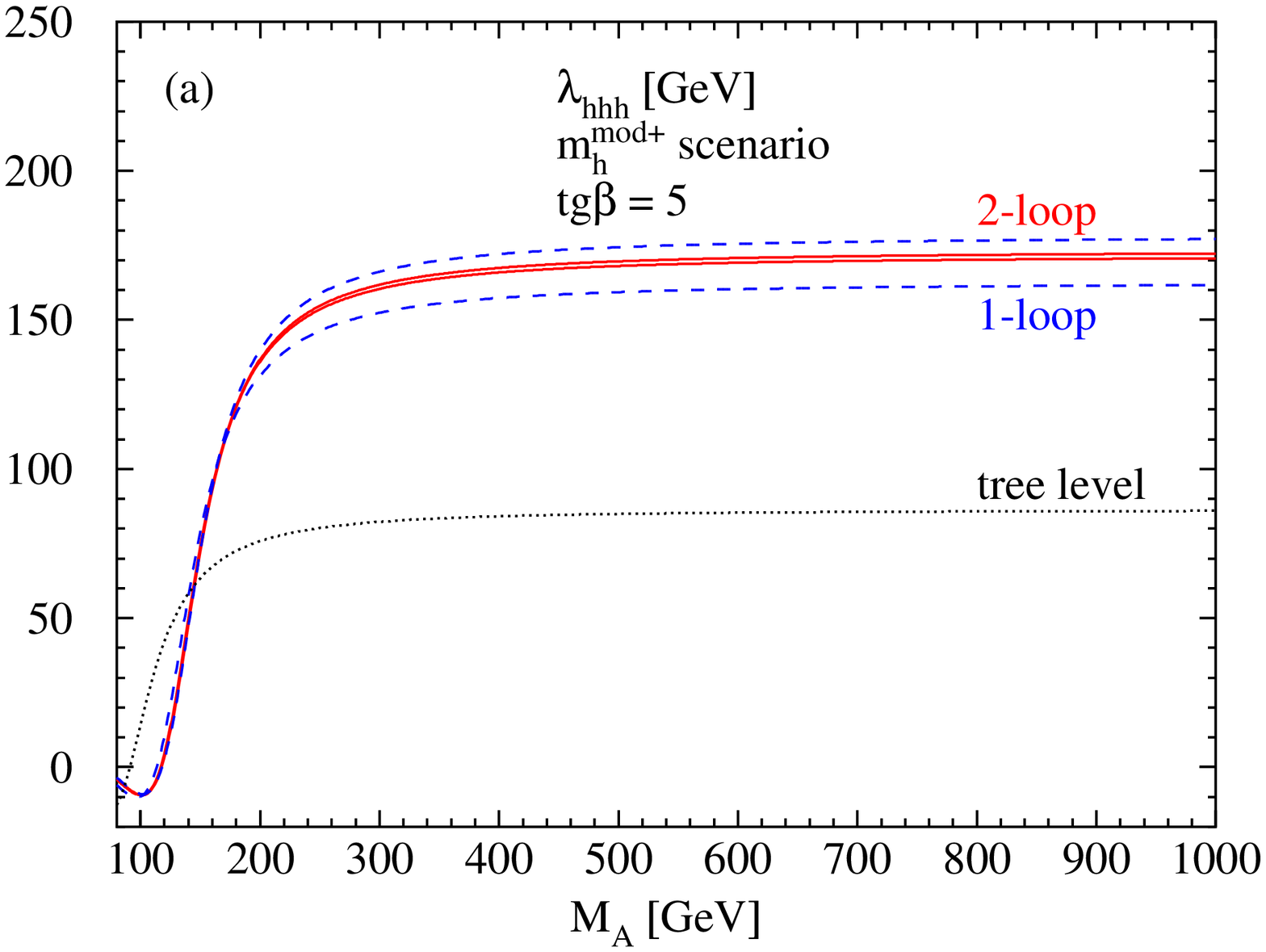,%
        bbllx=150pt,bblly=230pt,bburx=450pt,bbury=650pt,%
        scale=0.35,clip=}}
\put(30,0){\epsfig{file=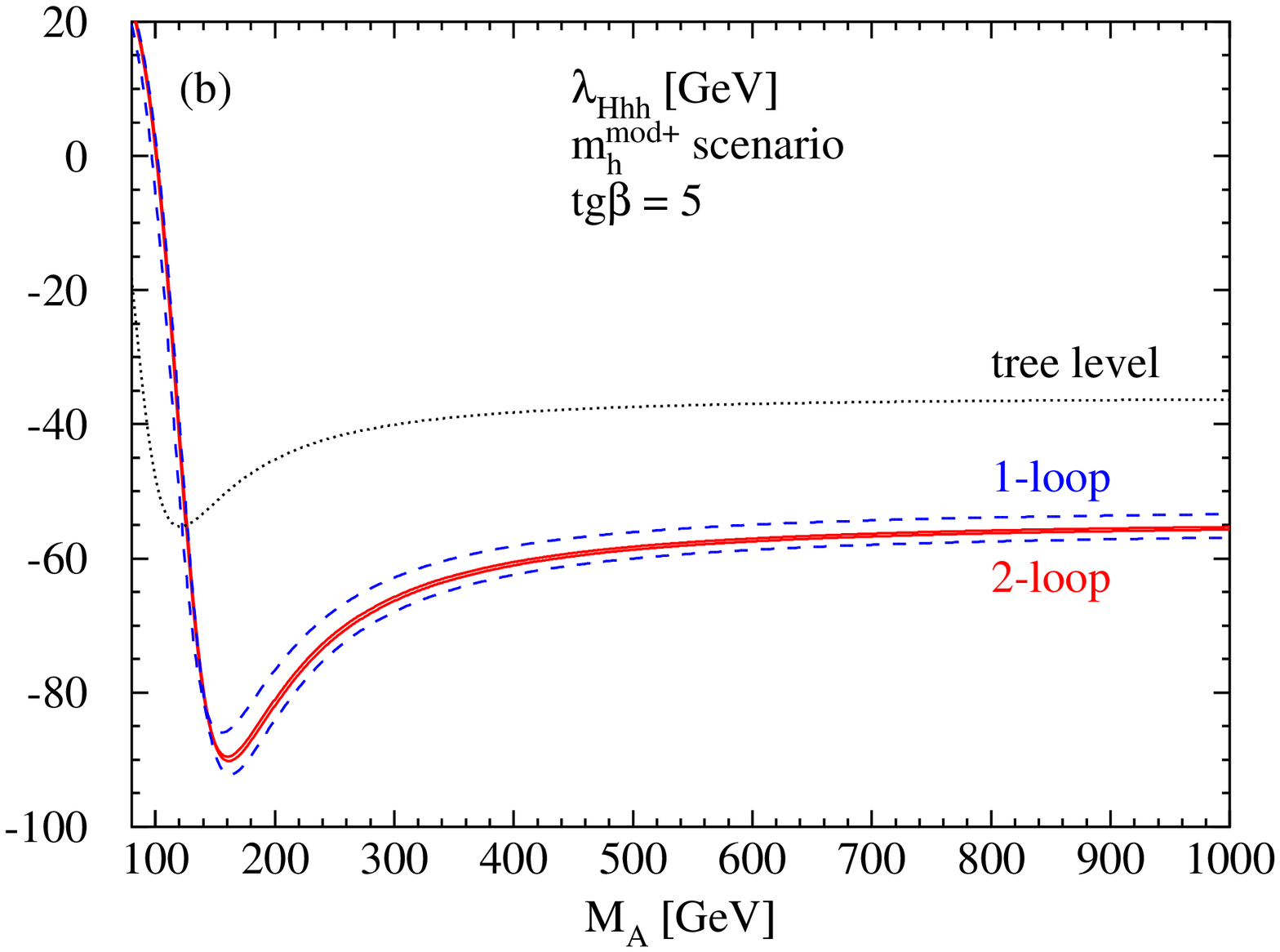,%
        bbllx=150pt,bblly=230pt,bburx=450pt,bbury=650pt,%
        scale=0.35,clip=}}
\end{picture}
\vspace*{-0.4cm}

\caption{\it \label{fg:bands} Trilinear Higgs
couplings $\lambda_{hhh}$ (a) and $\lambda_{Hhh}$ (b) at tree level, one- and
two-loop order in the $m_h^{mod+}$ scenario for $\tgb=5$. The bands
indicate the scale uncertainties within 1/3 and 3
times the central scale $M_{SUSY}/2$.}
\vspace*{-0.75cm}

\end{figure}

The scale dependences of the trilinear Higgs couplings $\lambda_{hhh}$
and $\lambda_{Hhh}$ are displayed in Fig.~\ref{fg:scale} at one- and
two-loop order. The central scale is chosen as half the SUSY scale,
i.e.~$\mu_R=M_{\tilde Q_{L/R}}/2 = 500$ GeV.  We obtain a significant
reduction of the scale dependence from ${\cal O}(10\%)$ at one-loop
order to the per-cent level at two-loop order and thus a large reduction
of the theoretical uncertainties. Moreover a broad maximum/minimum
develops at about the chosen central scale in contrast to the
monotonous scale dependences at one-loop order. In the ``$m_h^{mod+}$''
scenario the one-loop corrections are large and positive, increasing the
trilinear self-couplings by about a factor of 2. The two-loop
corrections amount to a few per cent for the central scale choices.
The strong reduction of the residual scale dependences is also visible in
Fig.~\ref{fg:bands}, which displays the trilinear Higgs couplings
$\lambda_{hhh}$ and $\lambda_{Hhh}$ as a function of the pseudoscalar
Higgs mass $M_A$. The one- and two-loop bands show the minimal and maximal
values of the Higgs couplings if the scale is varied between 1/3 and 3
times the central scale.

In summary, the significant scale dependence of ${\cal O}(10\%)$ of the
one-loop predictions for the trilinear MSSM Higgs self-couplings
requires the inclusion of two-loop corrections. For the corrected
trilinear and quartic Higgs couplings, we find a reduction of the scale
dependence to the per-cent level at ${\cal O}(\alpha_t\alpha_s)$. The
improved predictions for these couplings can thus be taken as a base for
experimental analyses at the LHC and the ILC.

\begin{acknowledgments}
We are indebted to P.\,Slavich for giving access to his PhD thesis and to
M.\,M\"uhlleitner and P.\,Zerwas for comments on the manuscript. This work
is supported in part by the Swiss National Science Foundation and by the
DFG SFB/TR9 ``Computational Particle Physics''.
\end{acknowledgments}

\end{document}